# CRITICAL FIELD OF MGB$_2$: CROSSOVER FROM CLEAN TO DIRTY REGIMES

M.Putti[a], V.Braccini[a], C.Ferdeghini[a], I.Pallecchi[a], A.S.Siri[a], F.Gatti[b], P.Manfrinetti[c], A.Palenzona[c].

[a] INFM-LAMIA, Dipartimento di Fisica, Università di Genova, via Dodecaneso 33, 16146 Genova Italy

[b] Dipartimento di Fisica, Università di Genova, via Dodecaneso 33, 16146 Genova Italy

[c] INFM-LAMIA, Dipartimento di Chimica e Chimica Industriale, via Dodecaneso 31, 16146 Genova, Italy

## ABSTRACT

We have studied the upper critical field, $B_{c2}$, in poly-crystalline MgB$_2$ samples in which disorder was varied in a controlled way to carry selectively π and σ bands from clean to dirty limit. We have found that the clean regime survives when π bands are dirty and σ bands are midway between clean and dirty. In this framework we can explain the anomalous behaviour of Al doped samples, in which $B_{c2}$ decreases as doping increases.

The two gap superconductivity, studied since the fifties, has found a very nice paradigm in MgB$_2$ which shows two well separated gaps associated with the two band sets which cross the Fermi level. The larger gap $\Delta_\sigma$ (≈7 meV) is related to the two-dimensional p-type σ bands, strongly coupled with optical boron modes, whereas the smaller gap $\Delta_\pi$ (≈2 meV) is related with the three-dimensional π bands.[1,2] π and σ bands have nearly equal density of states, so that both contribute significantly to the physical properties.

The upper critical field $B_{c2}$ was one of the first properties that emphasized the existence of two bands[3]. $B_{c2}$ is also anisotropic, the critical field parallel to the c-axis, $B_{c2,c}$, being lower than the perpendicular component, $B_{c2,ab}$. $B_{c2,ab}(0)$ values up to 50-60 T have been measured in thin films,[4,5] while the typical values reported for single crystals are of the order of 15-20 T.[6-8]

The critical field anisotropy $\gamma_{B_{c2}} = B_{c2,ab}/B_{c2,c}$ is temperature dependent and both its value and temperature dependence change from single crystals to thin films. In single crystals[7] $\gamma_{B_{c2}}$ shows the largest values at low temperature (6-7) and decreases down to 3-4 close to $T_c$, while in thin films $\gamma_{B_{c2}}$ has lower values (1.5-3.5) and is less dependent on temperature.[4,5] Many models have been proposed to explain this complex scenario, taking into account the presence of two bands with quite different intrinsic properties and various amounts of disorder.

Models of the mixed state in the clean limit[9] predict the coexistence of larger vortices related to the π bands ($\xi_{0_\pi} \approx 50$ nm) [10] and smaller vortices related to the σ bands ($\xi_{0_\sigma} \approx 12$ nm)[8]. At low temperature the critical fields are mainly related to the σ bands, due to their larger gap, and the anisotropy of critical fields in the clean limit, $\gamma_{B_{c2}^c}$, is predicted to be equal to the anisotropy of the effective mass tensor of σ bands, $\gamma_{B_{c2}^c}(0) \approx 6$; with increasing temperature, the thermal mixing with π-states suppresses $\gamma_{B_{c2}^c}$ down to 2.6 at $T_c$.[11] These outcomes agree very well with experimental data in clean samples.[6-8,12]

The transition from clean to dirty regime occurs when the electron mean free path becomes lower than the BCS coherence length; therefore, because of the larger values of $\xi_{0_\pi}$ with respect to $\xi_{0_\sigma}$, the π bands can be carried in dirty regime more easily than the σ bands. Recently some models which calculate the upper critical fields in the dirty limit have been proposed.[13,14] They have shown that the upper critical fields can show different values and shape, depending on whether the diffusivity in the π or in the σ bands prevails. Thus the simple correlation between the residual resistivity and $B_{c2}$ fails.[4] To examine closely the critical field behavior, a separate characterization of the mobility in the two bands becomes crucial.

In this letter we study the upper critical field of bulk samples to analyze the crossover from a clean regime (both bands clean), to one that is partly clean (σ bands clean and π bands dirty) and finally, to a dirty regime (both bands dirty). Starting from the cleanest samples, we introduced a controlled amount of disorder, carrying the two bands selectively from the clean to the dirty condition. The central point is to define which regime each band is in, so that the proper model for $B_{c2}$ is applied. We extract this information from thermal conductivity measurements, which allow the evaluation of the gap energies as well as the relaxation rates of each band.[15]

We selected five well characterized samples. They were prepared by direct synthesis from crystalline elements.[16] Two undoped samples were prepared from a stoichiometric mixture of crystalline Mg (99.999% purity) and crystalline boron from two different sources: isotopically enriched [11]B from Eagle-Picher (99.95% purity – UD1) and natural B from Alfa Aesar (-325 mesh, 99.97% purity – UD2). Two doped samples were prepared by using Alfa Aesar crystalline B and by substituting 5% (ALD1) and 10% (ALD2) Al on the Mg site. The fifth sample (IR) was obtained by neutron irradiation of a piece of UD1. The irradiation was carried out at the thermal neutron irradiation facility of the University of Pavia (LENA) with nominal fluence of $10^{18}$ neutron/cm$^2$.

The crystal defects are produced by neutron capture reactions in $^{10}$B, that are followed by α particle and $^{7}$Li nucleus emissions. The low $^{10}$B concentration in the sample (less then 0.5%) contains the self-shielding effect of neutrons into negligible values, providing an isotropic and homogeneous disorder.

All the samples were cut in the shape of a parallelepiped bar (~1×2×12 mm$^3$). Resistivity measurements were performed in a Quantum Design PPMS in magnetic field up to 9 T; thermal conductivity measurements were performed in a home-built cryostat working from 3 to 250 K.

X-ray powder patterns were obtained by a Guinier-Stoe camera; no extra peaks due to the presence of free Mg or spurious phases were detected, only traces of MgO due to the long exposition of powders in air. The doped compounds showed MgB$_2$ peaks shifted and slightly broadened, and the lattice parameters for ALD1 ($a$= 3.079(1) Å and $c$= 3.489(1) Å) and ALD2 ($a$= 3.077(1) Å and $c$= 3.483(1) Å) are in good agreement with other reports.[17,18]

In tab. I $T_c$, $\Delta T_c$, $\rho(40)$, the residual resistivity ratio (RRR) of the five samples are reported. The undoped compounds have optimal critical temperatures, sharp transitions and large RRR values. The Al doped samples present critical temperature which decreases with Al content, in good agreement with other reports,[17,18] and a slight broadening of the transition (mainly in ALD2). The increase in $\rho(40)$ can be related to the Al ions in the Mg planes, which act as scatterers mainly for the carriers in π bands.

As a consequence of the neutron irradiation, with respect to the pristine sample, the critical temperature of IR is lowered ($T_c$=36.3 K) and the residual resistivity is raised by a factor 30, but the transition still remains sharp, indicating that the neutrons penetrated the sample homogeneously.

In order to separate the contribution of the π and σ carriers to the conduction, thermal conductivity measurements were performed. The thermal conductivity in the superconducting state, is determined by quasi-particle excitations whose density depends exponentially on the energy gap. Thanks to the great difference between $\Delta_\pi$ and $\Delta_\sigma$, the contribution of π and σ excitations can be well discriminated. Actually, also phonons carry heat and only if their contribution is negligible reliable analysis can be carried out. We demonstrated that in very clean MgB$_2$ samples the phonon contribution can be neglected;[15] on the other hand, in large single crystals they give detectable contributions.[19]

Under the hypothesis of a negligible phonon contribution, the thermal conductivity in the superconducting state can be written as the sum of the π and σ thermal conductivities:[15]

$$\frac{\kappa^s(T)}{\kappa^n(T)} \approx [xg(T,\Delta_\pi)+(1-x)g(T,\Delta_\sigma)] \qquad (1)$$

where $\kappa^{s,n}(T)$ are the electron thermal conductivities in the superconducting and normal states, $g(T,\Delta_\alpha)$ ($\alpha=\sigma,\pi$) is a function that takes the quasi-particle condensation into account and the weights $x$ and ($1-x$) represent the energy fractions carried by the $\pi$ and $\sigma$ bands, respectively. If we assume that, at temperatures below $T_c$, the scattering with impurities dominates, the parameter $x$ can be written as $x = \rho_0 / \rho_{0_\pi}$ ($1-x = \rho_0 / \rho_{0_\sigma}$), where $\rho_0$ is the residual resistivity of the sample and $\rho_{0_\alpha}$ are those of $\pi$ and $\sigma$ bands. Given the $g$ function calculated in the BCS framework for scattering with impurities, eq. (1) can be used to fit the thermal conductivity data in the superconducting state with three free parameters $x, \Delta_\pi, \Delta_\sigma$. Finally, from $x$ and $\rho_0$ the relaxation rates of $\pi$ and $\sigma$ bands can be calculated:

$$\rho_{0_\pi} = \frac{\rho_0}{x} = \frac{\Gamma_\pi}{\omega_{p_\pi}^2 \varepsilon_0} \quad ; \quad \rho_{0_\sigma} = \frac{\rho_0}{1-x} = \frac{\Gamma_\sigma}{\omega_{p_\sigma}^2 \varepsilon_0} \tag{2}$$

where $\varepsilon_0$ is the dielectric constant, $\omega_{p_{\pi,\sigma}}$ are the plasma frequencies averaged in the three directions ($\omega_{p_\pi} = 6.226$ eV and $\omega_{p_\sigma} = 3.403$ eV)[20] and $\Gamma_\alpha$ are the scattering rates with impurities for the $\pi$ and $\sigma$ bands.

In fig.1 the thermal conductivity, $\kappa$, of the three samples UD1, UD2 and ALD1 is plotted: for these samples the above analysis can be carried out without ambiguity, while for ALD2 and IR the reduced electron contribution to $\kappa$ makes the phonon contribution no longer negligible. The dotted lines in fig. 1 represent $\kappa^p(T)$ obtained by fitting the experimental curves above $T_c$ with a generalized Wiedermann-Franz law.[15] The continuous lines are obtained by best fitting the experimental curves below $T_c$ with eq. (1), whose parameters are listed in table II. It is clear that the agreement between the theoretical and the experimental curves is very good apart from the lowest temperature region, where the phonons might give a non negligible contribution. The energy gaps derived from the fit are $\Delta_\sigma$=6.8–6.1 meV, $\Delta_\pi$=2.1–1.8 meV for undoped samples and $\Delta_\sigma$=5 meV, $\Delta_\pi$=1.6 meV for Al doped one. These values agree well with those obtained from a specific heat analysis[18] and confirm the strong reduction of the larger gap with Al doping.

Let us now consider the relaxation rates. The $x$ parameter, which is proportional to the mobility of the $\pi$ carriers, decreases from the cleanest ($x$=0.85) to the Al doped sample ($x$=0.6), indicating that in the latter the $\pi$ bands are more affected by disorder than the $\sigma$ bands. The relaxation rates calculated from eq. (3) show that, going from UD1 to ALD1, $\Gamma_\pi$ rises by a factor of 20, while $\Gamma_\sigma$ only by a factor of 5.

The next step is to compare the BCS coherence lengths ($\xi_{0_\sigma}, \xi_{0_\pi}$) with the relevant electron mean free paths ($l_\pi$, $l_\sigma$), which is equivalent to comparing the energy gaps with the relaxation rates: $\frac{\xi_{0_\alpha}}{l_\alpha} = \frac{\hbar v_{F_\alpha}/\pi \Delta_\alpha}{\hbar v_{F_\alpha}/\Gamma_\alpha} = \frac{\Gamma_\alpha}{\pi \Delta_\alpha}$. In table II the ratios $\xi_{0_\alpha}/l_\alpha$ are reported for the three samples. Each sample presents a different situation: UD1 has both clean bands ($\xi_{0_\sigma}/l_\sigma$ and $\xi_{0_\pi}/l_\pi < 1$); UD2 has clean σ bands ($\xi_{0_\sigma}/l_\sigma < 1$) and intermediate π bands ($\xi_{0_\pi}/l_\pi \sim 1$); ALD1 has intermediate σ bands ($\xi_{0_\sigma}/l_\sigma \sim 1$) and dirty π bands ($\xi_{0_\pi}/l_\pi \gg 1$). This analysis was not carried out for the samples ALD2 and IR, but some hypotheses can be made. ALD2 is very similar to ALD1, having the same kind of disorder localized on the Mg planes and nearly equal $\rho(40)$; therefore, a similar situation with intermediate σ bands and dirty π bands can be assumed. IR has large residual resistivity and disorder is homogeneously distributed: it is reasonable to assume that both bands are dirty.

The outlined scenario, summarized in table III, is quite complex and offers the possibility to study the evolution of the upper critical field when bands are selectively driven from clean to dirty regime.

In fig. 2 $B_{c2}$ is plotted as a function of the reduced temperature $t=T/T_c$ for all the samples. The curves $B_{c2}(T)$ were operatively defined at 90% of the transition. In polycrystalline samples this definition provides the $B_{c2,ab}$, corresponding to a path of connected grains with the ab plane parallel to the field. Fig. 2 illustrates that the $B_{c2}$ values can vary by more than a factor five in the five samples. $B_{c2}$ measured in the undoped samples nearly overlap, and assume the same values of $B_{c2,ab}$ measured in single crystals. In comparison with it, with increasing Al content, $B_{c2}$ strongly decreases while in the irradiated sample it increases. The behavior of Al doped samples is quite anomalous; in fact, unlike in common superconducting alloys, $B_{c2}$ decreases with increasing resistivity; this means that impurity scattering does not influence the critical field value as in the dirty limit. From the previous analysis it follows that these samples have dirty π bands and intermediate σ bands: we can still try to analyze them in a clean limit framework.

In clean limit the critical field parallel to the ab-planes is related to the σ bands:

$$B^c_{c2,ab}(T) = B^c_{c2,ab,\sigma}(T) = \frac{\Phi_0}{2\pi \xi_{0_\sigma}(T)^2} \gamma_{B^c_{ce}} \tag{3}$$

where $\gamma_{B_{c2}^c}$ is the anisotropy factor in the clean limit as defined by Miranovich et al..[11] At low temperatures $\gamma_{B_{c2}^c} \approx 6$ and decreases with temperature. Close to $T_c$ we can calculate the slope of eq. (3) assuming a BCS behavior for the coherence length:

$$\left.\frac{dB_{c2,ab}^c}{dT}\right|_{T_c} \approx -0.29 \frac{\Phi_0}{T_c} \frac{\gamma_{B_{c2}^c}(T_c)}{\xi_{0_\sigma}^2} \qquad (4)$$

where the term proportional to the anisotropy factor slope vanishes at $T_c$. $\xi_{0_\sigma}$ calculated from eq.(4) by inserting $\gamma_{B_{c2}^c}(T_c)=2.6$ [11] are summarized in table III for the four samples UD1, UD2, ALD1, ALD2. We find $\xi_{0_\sigma}=14\pm 2$ nm for the undoped samples in good agreement with the values estimated in single crystals.[8] Interestingly, by increasing the Al content, $\xi_{0_\sigma}$ rises up to $22\pm 3$ nm. Finally, the energy gap $\Delta_\sigma$ can be calculated from $\xi_{0_\sigma}$, once the in plane Fermi velocity of $\sigma$ bands ($v_{F_\sigma}=4.4\cdot 10^5$ m/s)[20] is introduced. The results of this calculation, indicated with $\Delta_\sigma^B$, are summarized in table III. In the same table the $\Delta_\sigma$ values obtained from thermal conductivity and specific heat measurements are reported as a comparison. The agreement between $\Delta_\sigma^B$ and $\Delta_\sigma$ is impressive, despite the approximations made. In particular, in the calculation we introduced the $\gamma_{B_{c2}^c}(T_c)$ and $v_{F_\sigma}$ values of the undoped compound also in the case of Al doped ones, but such overestimations compensate in the calculation.

In conclusion, our main result is that the strong suppression of $B_{c2}$ in the Al doped samples can be mainly ascribed to the reduction of the $\sigma$ gap. It is noteworthy that Al doping does not affect the $\sigma$ mean free path significantly, leaving $\sigma$ bands close to the clean regime; on the other hand, Al doping, changing electronic structure and increasing inter-band scattering rate, strongly suppresses $\Delta_\sigma$, as recently stated by specific heat measurements.[18] This picture is reminiscent of the superconductors with magnetic impurities. A low concentration of magnetic impurity does not affect the electron mean free path, but strongly suppresses the critical temperature and the energy gap; also in this case the critical field decreases.[21]

We analyze now the critical field for the irradiated sample, for which we assumed that both bands are in dirty limit. The fact the critical field of this sample is larger than in undoped samples strongly support this assumption. According to the model proposed by Gurevich,[13] the slope of $B_{c2,ab}$ close to $T_c$ in dirty is given by:

$$\left.\frac{dB_{c2,ab}^d}{dT}\right|_{T_c} \approx \frac{8\Phi_0 k_B}{\pi^2 \hbar} \frac{1}{a_1 D_\sigma / \gamma_{D_\sigma} + a_2 D_\pi} \tag{5}$$

where $a_1$=1.93 and $a_2$=0.07 are parameters related to the coupling constant, and $D_\pi = D_{\pi,c} \approx D_{\pi,ab}$ and $D_\sigma / \gamma_{D_\sigma}^2 = D_{\sigma,ab} / \gamma_{D_\sigma}^2 = D_{\sigma,c}$ are the diffusivities of the σ and π carriers in the c-direction. $\gamma_{D_\sigma}$ is the anisotropy of σ diffusivity and, within the dirty limit models,[12,13] it is nearly equal to the anisotropy of the critical fields $\gamma_{B_{c2}^d}(0)$. $\gamma_{B_{c2}^d}(0)$ can be estimated in thin films rather than in single crystals, and it is generally between 2 and 3.5.[4] We can fix $\gamma_{D_\sigma}$=3 and, considering that in the irradiated sample disorder is homogeneously distributed, we can assume $D_\sigma \sim D_\pi$. With this simplification eq. (5) correlates the critical field slope with the residual resistivity: for IR we find $D_\sigma = D_\pi = 1.24 \times 10^{-3}$ m$^2$s$^{-1}$. From this value the residual resistivities of the π and σ bands can be calculated from the relationships $(\rho_{0_\alpha})^{-1} = e^2 N_\alpha D_\alpha$, where $N_\sigma$=0.3 states/(eV·cell) and $N_\pi$= 0.4 states/(eV·cell) are the density of states. We obtain $\rho_{0_\sigma} \approx 50$ μΩcm, $\rho_{0_\pi} \approx 35$ μΩcm and $\rho_0$ calculated as the parallel of them comes out to be nearly 21 μΩcm, which is very close to the measured value $\rho(40)$=20 μΩcm. Our rough approximations make this agreement only indicative, but prove that in this case $B_{c2}$ is determined by scattering with impurities.

Finally, we have studied the upper critical field in MgB$_2$ samples in which disorder was varied in a controlled way and the crossover between the clean and dirty regime in each band has been monitored by means of the analysis of thermal conductivity data.

We observed that a clean limit picture, where $B_{c2}$ scales with the energy gap rather than with the relaxation rates, is quite robust and survives even when π bands are dirty and σ bands are midway between clean and dirty. Within this scenario we can explain the anomalous behaviour of Al doped samples in which $B_{c2}$ decreases as the Al content increases, although resistivity rises. In these samples only π bands are carried in the dirty limit; the decrease in $B_{c2}$ comes out straightforwardly from the suppression of the larger energy gap. When also σ bands become dirty, as it occurs in the irradiated sample, $B_{c2}$ increases, being determined by scattering mechanisms in π and σ bands.

This work is partially supported by I.F.N.M. through the PRA-UMBRA. The authors thank V.Ferrando, C.Tarantini and R.Vaglio for useful discussions.

**FIGURE CAPTIONS**

**Fig. 1**. $\kappa$ as a function of $T$ for UD1, UD2 and ALD1; the dotted lines are $\kappa^n(T)$ and continuous lines are $\kappa^s(T)$ obtained by best fitting the experimental curves above and below $T_c$, respectively (see the text).

**Fig. 2**. $B_{c2}$ as a function of $t=T/T_c$ .

**TABLE CAPTIONS**

**Tab.I:** $T_c$, $\Delta T_c$, $\rho(40)$ and RRR=$\rho(300)/\rho(40)$ of the five samples.

**Tab.II:** $x$, $\Delta_\sigma$ and $\Delta_\pi$ as obtained by fitting the thermal conductivity of the samples UD1, UD2 and ALD1; impurity scattering rates, $\Gamma_\sigma$ and $\Gamma_\pi$ obtained from eq. (2) and $\xi_{0_\alpha}/l_\alpha$ calculated as $\pi\Delta_\alpha/\Gamma_\alpha$.

**Tab.III:** $dB_{c2}/dT|_{T_c}$ estimated from the experimental curves; for UD1, UD2, ALD1 and ALD2 $\xi_{0\sigma}$ and $\Delta_\sigma^H$ are calculated from eq.(3); $\Delta_\sigma$ values are obtained from thermal conductivity and for ALD2 from specific heat (ref. 18).

**Table 1**

|      | $T_c$ K | $\Delta T_c$ K | $\rho(40)$ μΩcm | RRR |
|------|---------|----------------|-----------------|-----|
| **UD1**  | 38.7 | 0.2 | 0.6±0.05 | 15  |
| **UD2**  | 38.9 | 0.3 | 2.2±0.1  | 7   |
| **ALD1** | 35.7 | 0.5 | 6.7±0.3  | 2.7 |
| **ALD2** | 33.8 | 1.5 | 8.0±0.4  | 2.2 |
| **IR**   | 36.3 | 0.2 | 20±1     | 2.5 |

**Table II**

|      | $x$ | $\Delta_\sigma$ meV | $\Delta_\pi$ meV | $\Gamma_\sigma$ meV | $\Gamma_\pi$ meV | $\xi_{0\sigma}/l_\sigma$ | $\xi_{0\pi}/l_\pi$ |
|------|-----|------|------|------|------|------|------|
| **UD1**  | 0.85±0.05 | 6.8±1   | 1.8±0.1 | 4.5±1.3 | 3±1   | 0.20±0.06 | 0.5±0.2 |
| **UD2**  | 0.75±0.05 | 6.1±0.5 | 2.1±0.1 | 11±3    | 13±3  | 0.6±0.2   | 0.9±0.3 |
| **ALD1** | 0.60±0.05 | 5.0±0.5 | 1.6±0.2 | 24±7    | 53±15 | 1.5±0.6   | 11±4    |

**Table III**

|      |        | $dB_{c2}/dT|_{T_c}$ T/K | $\xi_{0\sigma}$ nm | $\Delta_\sigma^H$ meV | $\Delta_\sigma$ meV |
|------|--------|------------|-------|---------|---------|
| **UD1**  | σC πC | 0.19±0.03 | 14±2 | 6.4±1   | 6.8±1   |
| **UD2**  | σC πI | 0.19±0.03 | 14±2 | 6.4±1   | 6.1±0.5 |
| **ALD1** | σI πD | 0.13±0.02 | 18±1 | 5.1±0.7 | 5.0±0.5 |
| **ALD2** | σI πD | 0.10±0.01 | 22±3 | 4.3±0.5 | 4.4±0.2 |
| **IR**   | σD πD | 0.25±0.03 |      |         |         |

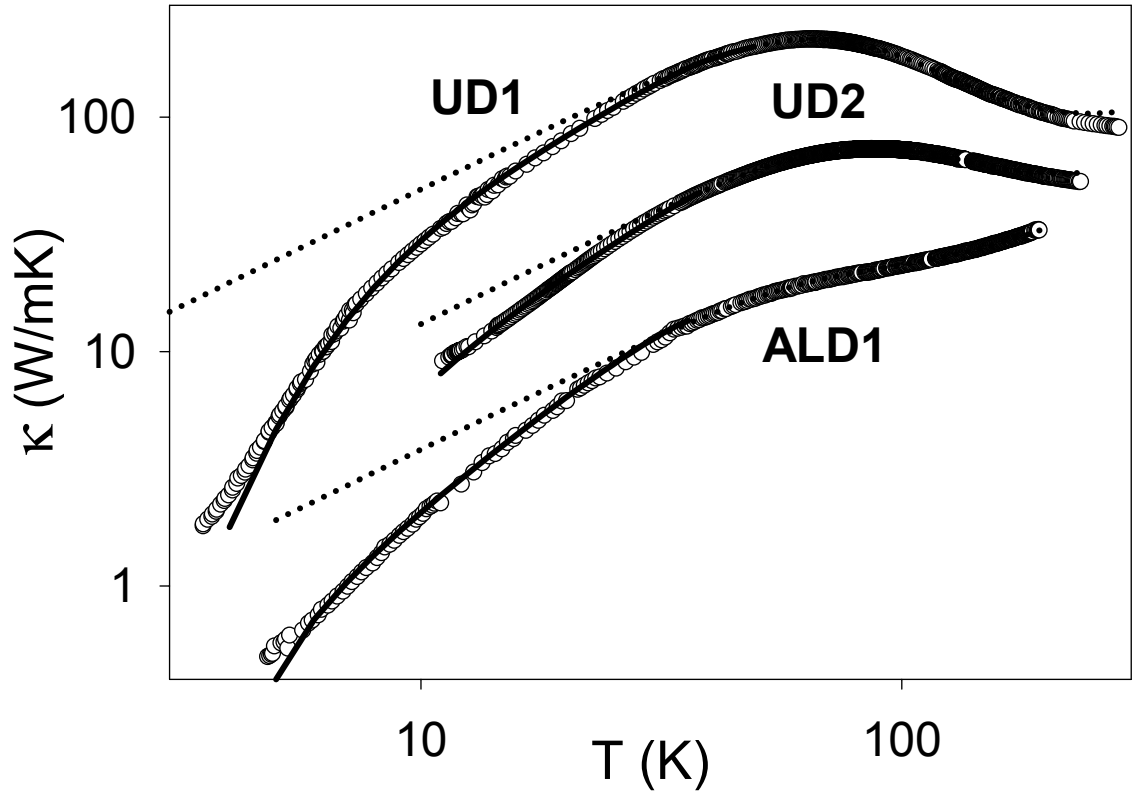

Figure 1

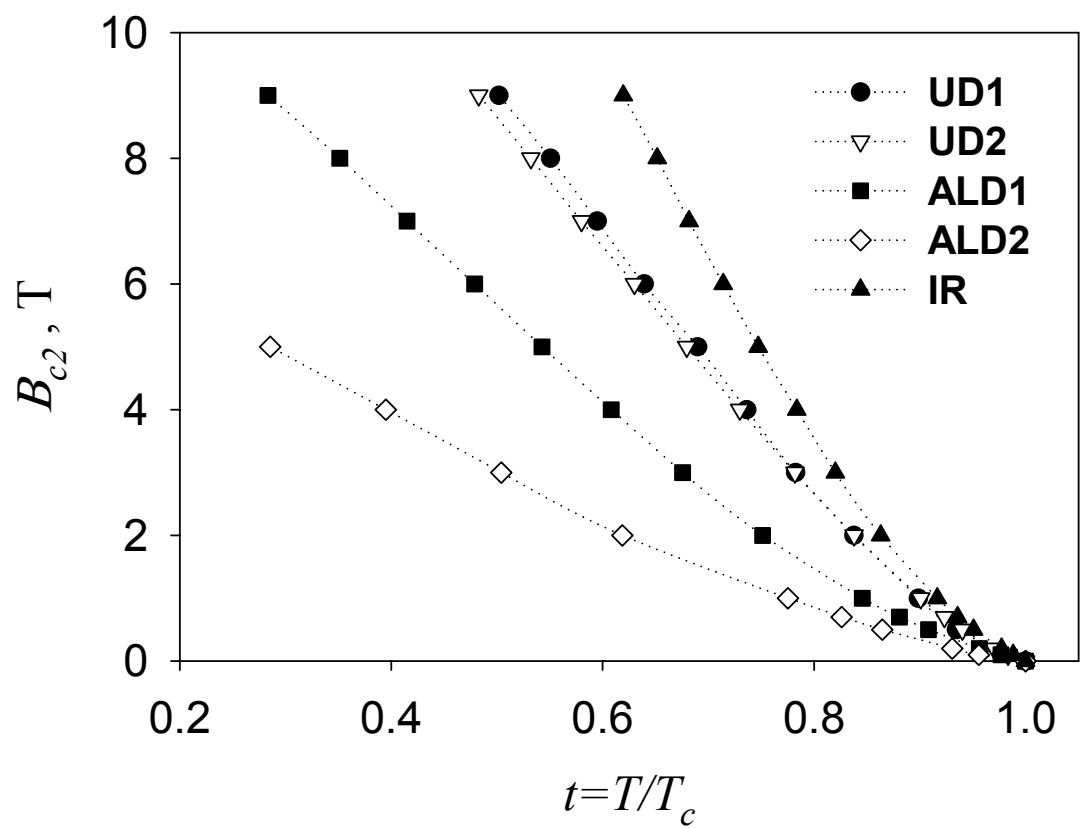

**Figure 2**